\documentclass[reprint,
superscriptaddress,
 amsmath,amssymb,
 aps,
 prapplied
]{revtex4-2}

\usepackage{graphicx}% Include figure files
\usepackage{dcolumn}% Align table columns on decimal point
\usepackage{bm}% bold math
\begin{document}

\preprint{APS/123-QED}

\title{Spin-orbit coupling and electron scattering in high-quality InSb$_{1-x}$As$_{x}$ quantum wells}

\author{S. Metti}
\affiliation{Elmore Family School of Electrical and Computer Engineering, Purdue University, West Lafayette, Indiana 47907, USA}
\affiliation{Birck Nanotechnology Center, Purdue University, West Lafayette, Indiana 47907, USA}

\author{C. Thomas}
\affiliation{Department of Physics and Astronomy, Purdue University, West Lafayette, Indiana 47907, USA} 
\affiliation{Birck Nanotechnology Center, Purdue University, West Lafayette, Indiana 47907, USA}
\author{D. Xiao}
\affiliation{Department of Physics and Astronomy, Purdue University, West Lafayette, Indiana 47907, USA} 
\affiliation{Birck Nanotechnology Center, Purdue University, West Lafayette, Indiana 47907, USA}
\author{M. J. Manfra$^{\dagger}$}
\affiliation{Department of Physics and Astronomy, Purdue University, West Lafayette, Indiana 47907, USA} 
\affiliation{Birck Nanotechnology Center, Purdue University,  West Lafayette, Indiana 47907, USA}
\affiliation{Elmore Family School of Electrical and Computer Engineering, Purdue University, West Lafayette, Indiana 47907, USA}
\affiliation{School of Materials Engineering, Purdue University, West Lafayette, Indiana 47907, USA}
\affiliation{Microsoft Quantum Lab West Lafayette, West Lafayette, Indiana 47907, USA}

\date{\today}% It is always \today, today,

\begin{abstract}
 InSb$_{1-x}$As$_{x}$ is a promising material system for exploration of topological superconductivity in hybrid superconductor/semiconductor devices due to large effective $g$-factor and enhanced spin-orbit coupling when compared to binary InSb and InAs. Much remains to be understood concerning the fundamental properties of the two-dimensional electron gas (2DEG) in InSbAs quantum wells. We report on the electrical properties of a series of 30 nm InSb$_{1-x}$As$_{x}$ quantum wells grown 40 nm below the surface with three different arsenic mole fractions, $x=$ 0.05, 0.13, and 0.19. The dependencies of mobility on 2DEG density and arsenic mole fraction are analyzed. For the $x = 0.05$ sample, the 2DEG displays a peak mobility $\mu = 2.4$ $ \times$ $ 10^5$ cm$^{2}$/{Vs} at a density of $n = 2.5$ $ \times$ $ 10^{11}$ cm$^{-2}$. High mobility, small effective mass, and strong spin-orbit coupling result in beating in the Shubnikov de Haas oscillations at low magnetic field. Fourier analysis of the Shubnikov de Haas oscillations facilitates extraction of the Rashba spin-orbit parameter $\alpha$ as a function of 2DEG density and quantum well mole fraction. For $x = 0.19$ at $n = 3.1 $ $ \times$ $ 10^{11}$ cm $^{-2}$, $\alpha \approx 300$ meV \text{\AA}, among the highest reported values in III-V materials. 
\end{abstract}

\maketitle

\section{\label{sec:level1}Introduction}
Small band gap zinc-blende III-V semiconductors (such as InAs and InSb) are the subject of renewed interest as they may be utilized as a platform to explore topological superconductivity. Strong spin-orbit coupling (SOC) and proximity-induced superconductivity from a parent s-wave superconductor are key ingredients needed to realize a topological phase \cite{Lutchyn.2010,Oreg.2010, Shabani.2016,Kjaergaard.2016, Drachmann.2017,Ke2019, fornieri2019evidence, mayer2020superconducting, Dartiailh.2021, Dartiailh.2021an9,Moehle2021}. Gate-controlled SOC is of fundamental and practical interest. The ability to externally manipulate the strength of SOC in a 2DEG facilitates realization of spin dependent devices, such as the spin field effect transistor \cite{datta1990electronic}.

The ternary compound InSb$_{1-x}$As$_{x}$ is appealing compared to InSb and InAs because it is predicted to have a higher Rashba SOC parameter $\alpha$ \cite{Berolo, Bouarissa, Winkler2016, mayer2020superconducting}. The increase in SOC with arsenic mole fraction in InSb$_{1-x}$As$_{x}$ has been measured in nanowires \cite{Sestoft2018} and recently in a surface 2DEG \cite{Moehle2021}. Here, we focus on 30 nm quantum wells buried under a 40 nm In$_{0.88}$Al$_{0.12}$Sb top barrier. Arsenic mole fractions of $x$ = 0.05, $x$ = 0.13, $x$ = 0.19 are used in the quantum well. Our buried quantum well design, compared to a surface layer, allows for more straightforward comparisons of the Rashba parameter between samples at nearly equal electric field and 2DEG density, as well as identification of electron scattering mechanisms over a broader range of 2DEG density.
\break
\indent Spin-orbit coupling and the resultant zero-field spin-split bands arises from broken inversion symmetry. Inversion asymmetry may be microscopic bulk inversion asymmetry (BIA) or structural inversion asymmetry (SIA). Dresselhaus SOC is associated with the lack of inversion symmetry in the crystal lattice, e.g. zinc-blende semiconductors \cite{dresselhaus1955spin}. The Rashba mechanism is due to asymmetry in the macroscopic potential induced by an external gate and/or an asymmetric heterostructure design \cite{bychkov1984oscillatory}. In heterostructures built with small bandgap III-V zinc-blende semiconductors, the Rashba coupling is usually the dominant $\textit{k}$-linear SOC term. Values for the Rashba parameter $\alpha$ for InSb and InAs quantum wells have been reported. Hatke et al. \cite{Hatke2017} reported $\alpha \approx$ 120 meV$ \text{\AA}$ in deep InAs quantum wells at  $n \approx$ 3 $\times$ 10$^{11}$ cm$^{-2}$; in modulation-doped InSb quantum wells Gilbertson et al. \cite{Gilbertson} extracted $\alpha$ between 130-160 meV$ \text{\AA}$ for a density range of  $n \approx$ 3.15-3.30 $\times$ 10$^{11}$ cm$^{-2}$.
We report Rashba coupling of $\alpha \approx$ 300 meV$ \text{\AA}$ in a InSb$_{0.81}$As$_{0.19}$ quantum well, among the highest measured in III-V semiconductors \cite{Nitta1997, Luo1990, Heida1998, Gilbertson, Kim2010, Kallaher2010,Lee2011, Shojaei2016, Hatke2017, lei2021high}.

\begin{table*}[t]
\centering
\caption{\label{tab:table2} Arsenic mole fraction, 2DEG density at zero gate bias, mobility at $n = 2.5 $ $\times$ $ 10^{11}$ cm$^{-2}$, and effective mass extracted by temperature dependent SdH oscillations for the samples used in this study. Mobility is measured at $T$ = 0.3 K}
\begin{ruledtabular}
\begin{tabular}{ccccc}{width=textwidth}
 & As mole fraction ($x$)
 & $n$ at V$_g$ = 0 V ($10^{11}$ cm$^{-2}$)  
 & ${\mu}$ at $n = 2.5 \times 10^{11}$cm$^{-2}$ (cm$^{2}$/Vs)
 & m* (m$_e$)\\
\hline
Sample A & 0.05 & 0.6  & 2.4 $\times$ $10^5$ & 0.0178 $\pm$ 0.0004  \\
Sample B & 0.13 & 1.4  & 2.0 $\times$ $10^5$ & 0.0142 $\pm$ 0.0002   \\
Sample C & 0.19 & 2.3  & 1.4 $\times$ $10^5$ & 0.0139 $\pm$ 0.0002   \\

\end{tabular}
\end{ruledtabular}
\label{table:info}
\end{table*}

\section{\label{exp}Experimental Details}
\begin{figure}[ht]
    \includegraphics[width=\columnwidth]{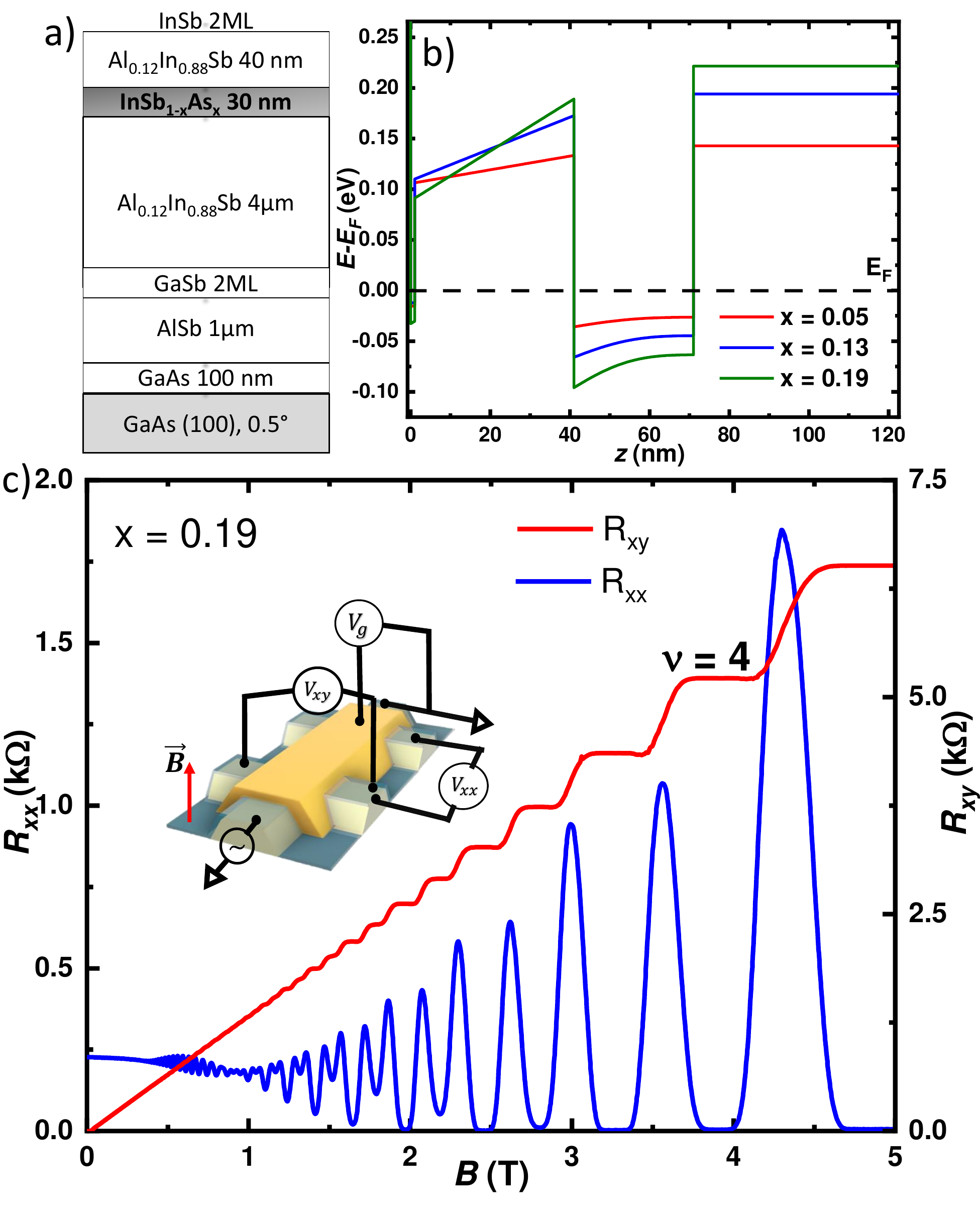}
    \caption{a) Schematic of the heterostructure used in this experiment. Arsenic mole fractions are $x$ = 0.05, 0.13 and 0.19 b) Band edge profile for $x$ = 0.05, 0.13 and 0.19 at $V_{g}$ = 0 V.  c) Magnetotransport measurement at $T$ = 0.3 K of Sample C at $V_{g}$ = 0.65 V with  $n =  4.6$ $\times$ 10$^{11}$ cm${^{-2}}$ and $\mu$ = 1.13 $\times$ 10$^{5}$  cm$^{2}$/Vs.}
    
    \label{Fig:fig1}
\end{figure}

We examine three heterostructures grown by molecular beam epitaxy (MBE) on GaAs (001) substrates with 0.5${^\circ}$ miscut towards (111)B. Three different arsenic mole fractions are used in the InSb$_{1-x}$As$_{x}$ quantum well: $x$ = 0.05, 0.13, and 0.19, labeled Sample A, B, and C respectively (see Table \ref{table:info}). To mitigate lattice-mismatch between the GaAs substrate and InSbAs quantum well and AlInSb barriers, the heterostructure (Fig. \ref{Fig:fig1}a) is built upon a relaxed buffer layer developed in previous work \cite{Ke2019}. The InSb$_{1-x}$As$_{x}$ quantum well is grown by keeping the In, Sb and As shutters open simultaneously; the arsenic mole fraction is determined by the substrate temperature and the Sb/As flux ratio.

Using a self-consistent Schr\"{o}dinger-Poisson (SP) simulator (NEXTNANO$^3$ \cite{Nextnano}) we calculate the band edge profile for the devices used in our experiments under changing gate potentials. As highlighted in Fig. \ref{Fig:fig1}b, in comparison to undoped InSb quantum wells \cite{kulesh2020quantum,lei2021high}, these undoped InSbAs samples have a conductive 2DEG channel in the quantum well at zero gate bias; higher carrier density at $V_g$ = 0 V is obtained with increasing arsenic mole fraction (see Table \ref{tab:table2}) \cite{Suchalkin2016,Moehle2021}.

 Insulated-gate Hall bars were fabricated to study electrical transport properties. Mesas were defined with a citric acid etch. The Hall bars typically are 75 $\mu$m wide and 1 mm long. The Ohmic probes are separated by 150 $\mu$m along the mesa edge. Before depositing the Ohmic metal stack an in-situ argon ion mill with a voltage of 250 V, a beam current of 8 mA, and an accelerating voltage of 50 V was performed to etch the top barrier such that the contact metal stack is deposited directly at the InSbAs quantum well. Secondary ion mass spectroscopy was used to determine the position of the quantum well. This milling process was followed by deposition of 20 nm Ti and 180 nm Au for the Ohmic contacts. Next 40 nm of Al$_2$O$_3$ was deposited as a dielectric via atomic layer deposition at 100 $^{\circ}$C. Finally, the gate electrodes were fabricated with the deposition of 20 nm of Ti and 150 nm Au.  

 Magnetotransport measurements were performed at temperatures between $T$ = 0.3 K and $T$ = 10 mK, unless otherwise specified. Higher temperature measurements were required for effective mass determination and investigation of magneto-intersubband scattering. In Fig. \ref{Fig:fig1}c we show representative high magnetic field data for Sample C.  Well-defined integer quantum Hall states with vanishing longitudinal resistance and quantized Hall resistance are evident. The density calculated from the Hall slope matches the density extracted from analysis of Shubniokov de Haas (SdH) oscillations, indicative of a single conducting channel.
We measured mobility as a function of 2DEG density for the three wafers at $T$ = 0.3 K to identify the dominant scattering mechanisms and to observe changes that may occur as the arsenic mole fraction is varied between samples. For determination of SOC parameters, our measurements focused on high resolution low magnetic field ($B\leq 1$ T) measurements of the longitudinal resistance oscillations.

\section{\label{results}Results and Discussion}

\subsection{\label{scatt}Electron scattering in ternary InSbAs quantum wells}
Scattering of 2D electrons has been thoroughly investigated in the binary compounds InSb \cite{Chung1999,Yi2015, A.MGilbertson2010,Lehner2018, lei2021high} and InAs \cite{Shabani2014, Hatke2017}. It was determined that at cryogenic temperatures mobility is primarily limited by long-range Coulomb scattering from remote and background impurities for 2DEG densities below 1 $\times$ 10$^{12}$ cm$^{-2}$. Here, we investigate mobility in ternary InSbAs quantum wells. A ternary quantum well may have additional sources of scattering. Anion atoms (Sb, As) will be randomly distributed based on the available sublattice sites, generating short-range disorder \cite{harrison1976alloy}. Alloy disorder and interface roughness scattering are sources of short-range scattering that may lower mobility and modify the functional dependence of mobility on 2DEG density in InSbAs channels \cite{chin1991electron,DasSarma2013}.

\begin{figure}[t]
    \includegraphics[width=\columnwidth]{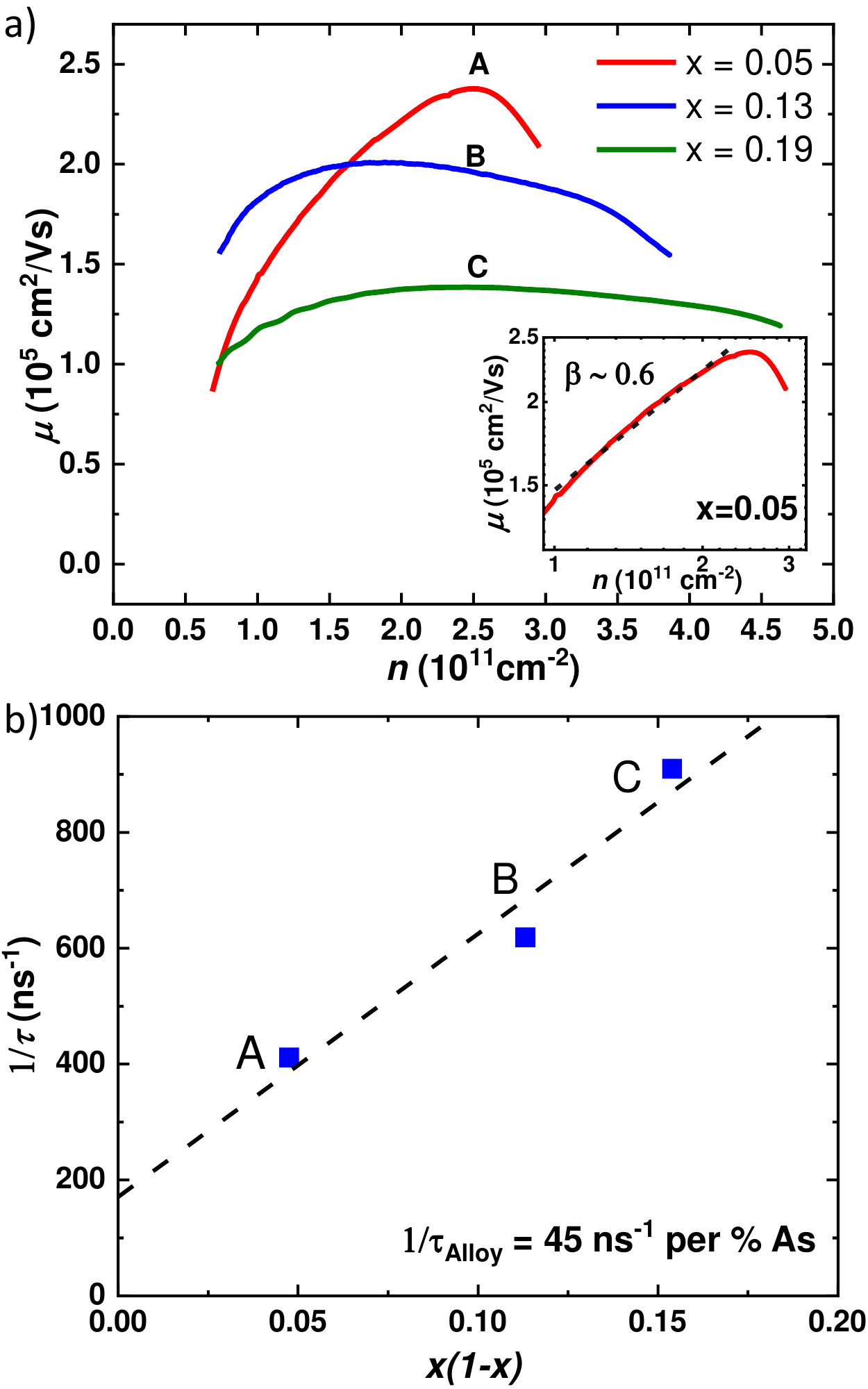}
    \caption{a) $\mu$ vs. 2DEG density measured at $T$ = 0.3 K for Samples A, B and C. Inset: $\mu$ vs. 2DEG density plotted in a log-log scale for Sample A and linear fit (dashed line) for extraction of $\beta$. b) Total scattering rate 1/$\tau$ vs $x(1-x)$ at $T$ = 0.3 K and $n$ = $2.5 \times 10^{11}$cm$^{-2}$, where \textbf{x} is the arsenic mole fraction. The dashed line yields 1/$\tau_{\text{alloy}}$ = 45 ns$^{-1}$ per $\%$ As. }
    
    \label{Fig:fig2}
\end{figure}

Mobility ($\mu$) vs. 2DEG density measured at $T$=0.3 K for different arsenic mole fractions is displayed in Fig. \ref{Fig:fig2}a. As the arsenic mole fraction increases from Sample A to Sample C, a decrease in peak mobility is observed. In addition, a change in the functional dependence of $\mu$ vs. 2DEG density is evident.  For Sample A at low density, mobility is a strongly increasing function of density, typical of charged impurity scattering. The dependence may be approximated by $\mu \propto n^{\beta}$ over the limited density range of $n = 1-2.2 \times 10^{11}$cm$^{-2}$, yielding an exponent of $\beta$ = 0.6, as seen in the inset of Fig. \ref{Fig:fig2}a with data plotted on a log-log scale. $\beta$ = 0.6 indicates that long-range charge disorder dominates scattering in this density regime, specifically Coulomb disorder near the quantum well \cite{DasSarma2013}. In addition to point defects randomly distributed in the vicinity of the quantum well, the semiconductor/dielectric interface is a known source of fixed surface charge. In our structures, this surface is 40 nm away from the quantum well. Below $n = 1 \times 10^{11}$cm$^{-2}$, the slope in the $\mu$ vs. $n$ data increases, suggestive of a transition to a localized regime at lowest density \cite{DasSarma2013}. 

We note that at a fixed density of 2.5 $\times10^{11}$ cm$^{-2}$, a decrease (increase) in mobility (scattering rate) with increasing arsenic mole fraction is observed, where $\mu=e\tau/m^*$. Mobility peaks at 2.4 $\times$ $10^5$ cm$^{2}$/Vs for Sample A, followed by Sample B and Sample C with mobility of 2.0 $\times$ $10^5$ cm$^{2}$/Vs and 1.4 $\times$ 10$^5$ cm$^{2}$/Vs at the same 2DEG density (Table \ref{table:info}). A recent work from Lei $et$ $al.$\cite{lei2021high} reports mobility of 2.6 $\times$ 10$^5$ cm$^{2}$/Vs at density 2.4 $\times$ $10^{11}$ cm$^{-2}$ in an undoped InSb quantum well, a value similar to that seen in Sample A, suggesting that alloy disorder scattering is not yet dominant at $x=0.05$.

For Sample B and Sample C, the weaker dependence of mobility on 2DEG density suggests that short-range scattering limits mobility at increased arsenic mole fraction \cite{DasSarma2013}. The progression of $\mu$ vs. $n$ from Sample A to B and C indicates that alloy disorder significantly impacts mobility at mole fractions above $x=0.05$. 

According to Matthiessen's rule, the total scattering rate ($1/\tau$) is determined by summing the scattering rates due to all independent scattering mechanisms. At fixed measurement temperature $T$ = 0.3 K, we assume that 1/$\tau = 1/\tau_\text{other} + 1/\tau_\text{alloy}$, where $1/\tau_\text{other}$ includes contributions for all scattering mechanisms other than alloy disorder scattering. The alloy scattering rate should be linear with $ m^{*} [\delta V]^2 x(1-x)$, where $x$ is the As mole fraction, $m^*$ is the effective mass and $[\delta V]^2$ is the alloy scattering potential \cite{bastard1990wave}. 

In order to assess the increase in scattering associated with the change in arsenic mole fraction we plot the total scattering rate ($1/\tau$) vs $x(1-x)$ at $T$ = 0.3 K and $n =  2.5 \times 10^{11}$ cm$^{-2}$ as seen in Fig. \ref {Fig:fig2}b. The linear dependence of the total scattering rate with $x(1-x)$ yields a slope of 1/$\tau$ = 45 ns$^{-1}$ per $\%$ As, which quantifies the impact of alloy scattering as we increase arsenic mole fraction.

To the best of our knowledge, the alloy scattering rate in InSbAs quantum wells has not been reported previously. Therefore, we compare the extracted alloy scattering rate of 1/$\tau_\text{alloy}$ = 45 ns$^{-1}$ per $\%$ As with the alloy scattering rate of $1/\tau_{\text{alloy}}=24$ ns$^{-1}$ per $\%$ As extracted by Gardner \textit{et al.} in the more widely studied Al$_{x}$Ga$_{1-x}$As ternary \cite{gardner2013growth}. The rate measured in InSb$_{1-x}$As$_{x}$ quantum wells is comparable to Al$_{x}$Ga$_{1-x}$As. $1/\tau_\text{alloy}$ is expected to be proportional to $[\delta V]^2$ and  m$^*$. The alloy scattering potential for an InSb-InAs alloy is theoretically predicted to be 0.82 eV compared to 0.12 eV for GaAs-AlAs alloy \cite{Ferry1978} while $m^*$ is $\approx$ 0.010-0.019m$_e$ for InSb$_{1-x}$As$_{x}$ ternary compared to 0.067m$_e$ for  Al$_{x}$Ga$_{1-x}$As \cite{gardner2013growth}. Therefore, the alloy scattering rate for InSb$_{1-x}$As$_{x}$ is within a factor of 2 of Al$_{x}$Ga$_{1-x}$As, despite the significantly higher alloy scattering potential, due to the compensatory impact of the lower effective mass. 

We performed temperature dependent magnetotransport measurements to extract the values of the effective mass for all samples at a density of $n \approx$ 2 $\times$ 10$^{11}$ cm${^{-2}}$ \cite{ihn2009}. Theory and experiments have shown that the ternary InSb$_{1-x}$As$_{x}$ has a lower effective mass with increasing arsenic mole fraction \cite{Bouarissa, Berolo,Suchalkin2016,Moehle2021}. Our analysis yielded results consistent with this trend, as displayed in Table \ref{table:info}.

\begin{figure}[hb]
    \includegraphics[width=\columnwidth]{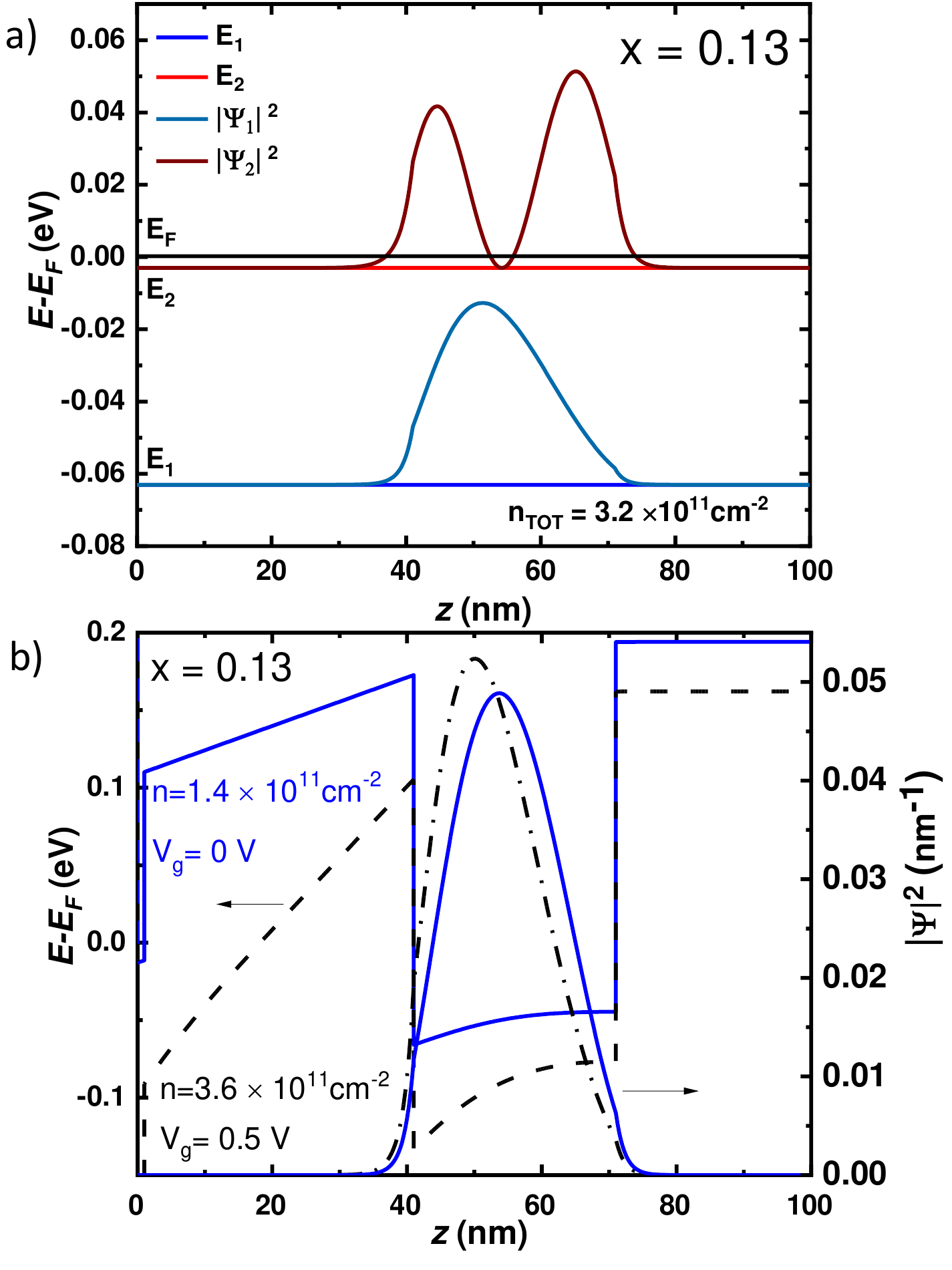}
    \caption{a) Self-consistent calculation of energy eigenvalues and wavefunctions of first and second subbands, when total density, $n = 3.2$ $\times$ 10$^{11}$ cm${^{-2}}$. b) Self-consistent calculation of band diagram and wavefunction of the first subband at V$_{g}$ = 0 V, with density $n = 1.4 $ $ \times$ 10$^{11}$ cm${^{-2}}$ (Solid line) and at V$_{g}$ = 0.5 V with density $n$ = 3.6 $\times$ 10$^{11}$ cm${^{-2}}$ (Dashed line). As the gate voltage increases, we observe enhanced band bending and asymmetry of the wavefunction within the quantum well. }
    
    \label{Fig:fig3}
\end{figure} 

Finally we note a decrease in mobility for all samples above $n \approx 2.5 \times 10^{11}$ cm$^{-2}$, as seen Fig. \ref{Fig:fig2}a.
Simulations point to the occupation of the second subband at a similar density for all samples, $ n \approx$ 3 $\times$ 10$^{11}$ cm$^{-2}$, as seen in Fig. \ref{Fig:fig3}a. The drop in mobility in our data may be attributed to the proximity of the Fermi level to the bottom of the second subband, allowing a channel for intersubband scattering \cite{Fletcher}. Another possible contribution is interface roughness scattering, since the electron wavefuncion is pulled closer to the top barrier with higher gate voltage, as seen in the simulation shown in Fig. \ref{Fig:fig3}b. 

\subsection{\label{soc}Analysis of spin-orbit coupling in InSbAs quantum wells}
Due to high mobility and small effective mass we are able to resolve beating in the oscillatory longitudinal resistance at low magnetic field. Beating may be driven by several mechanisms, including density inhomogeneity \cite{Brosig}, magneto-intersubband scattering \cite{leadley1992,Raikh,Sander1998,Rowe2013} or zero-field spin splitting arising from strong spin-orbit coupling \cite{Luo1988}. Magnetotransport measurements taken with different contact configurations on our devices indicate that beating does not result from density inhomogeneity. Intersubband scattering may occur when the chemical potential is proximal to the second electric subband. Landau level crossing enhances magneto-intersubband scattering at magnetic field given by $B_{MIS} = ({\Delta} E_{1,2}m^{*})/ (e\hbar)$ \cite{leadley1992, Raikh, Sander1998,Rowe2013}. As noted previously, simulations indicate the occupation of the second subband is possible at high 2DEG density, therefore we have measured temperature-dependent magnetotransport to rule out magneto-intersubband scattering.

We first note that the magnetotransport does not show signs of significant parasitic conduction. As shown in Fig. \ref{Fig:fig1}b, clear quantum Hall plateaus are resolved with the longitudinal resistance vanishing at 1.8 T. Furthermore, the densities extracted from the Hall slope and from the SdH oscillations minima are equal. This suggests minimal contribution from parallel conduction channels in the investigated range of density, with only the lowest energy subband contributing significantly to transport. 
 
\begin{figure}[t]
    \includegraphics[ scale=0.65]{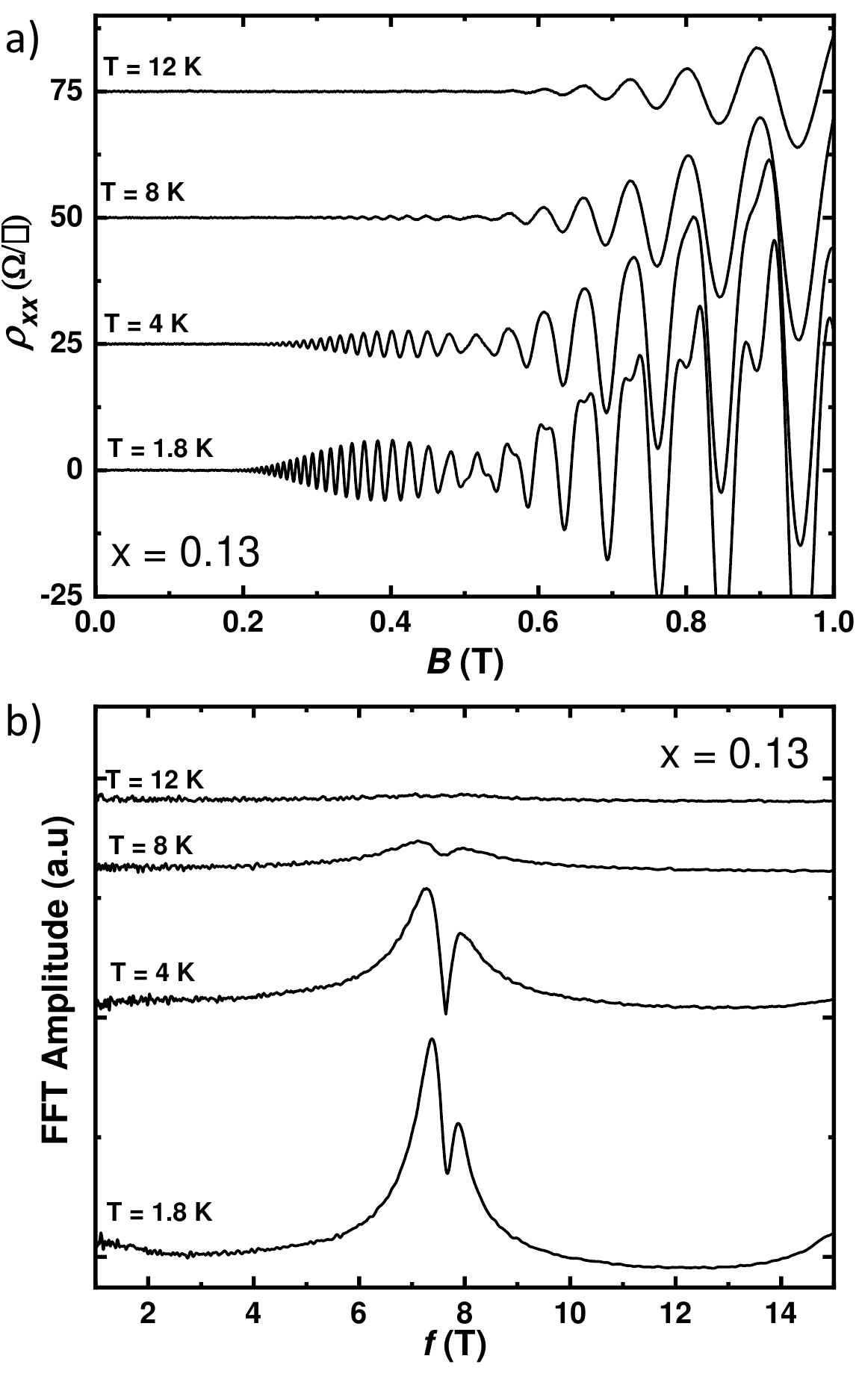}
    \caption{ a) Magnetotransport as a function of temperature for $x$ = 0.13 at density of  $n = 3.6 \times$ 10$^{11}$ cm${^{-2}}$. b) Fast Fourier transforms for the data shown in a). The damping of the FFT peaks as a function of temperature indicates that the peaks are generated by SdH oscillations and not by magneto-intersubband scattering.}
    \label{Fig:fig4}
\end{figure}

 Magneto-intersubband scattering occurs when the Landau levels of two bands cross, causing oscillations in the magnetotransport at frequency $B_{MIS} = ({\Delta} E_{1,2}m^{*})/ (e\hbar)$. If this value is close to the frequency of the underlying SdH oscillations, beating may result. To understand if this mechanism was active in our devices, measurements of the longitudinal resistance were performed over a broad range of temperature. Magneto-intersubband scattering is relatively insensitive to changes in temperature, whereas damping of the SdH oscillations with temperature occurs rapidly as $A(T) \propto x$/sinh$(x)$ with $x=(2{\pi}^2 k_{b}T/ {\hbar}{\omega}_{c})$ where ${\omega}_{c}=eB/{m}^*$ \cite{Rowe2013,Sander1998,Shojaei2016}.
We performed temperature dependent measurements for Sample B and Sample C at the highest achievable density. We observe rapid damping of the oscillations (Fig. \ref{Fig:fig4}a) with increased temperature, also reflected in the fast Fourier transforms (FFT). The amplitudes of the two frequency peaks clearly decrease with increasing temperature (Fig. \ref{Fig:fig4}b), indicating that their origin is SdH oscillations rather than magneto-intersubband scattering.

\subsection{\label{rashba} Rashba spin-orbit coupling}

\begin{figure}[t]
    \includegraphics[width=\columnwidth,scale=0.6]{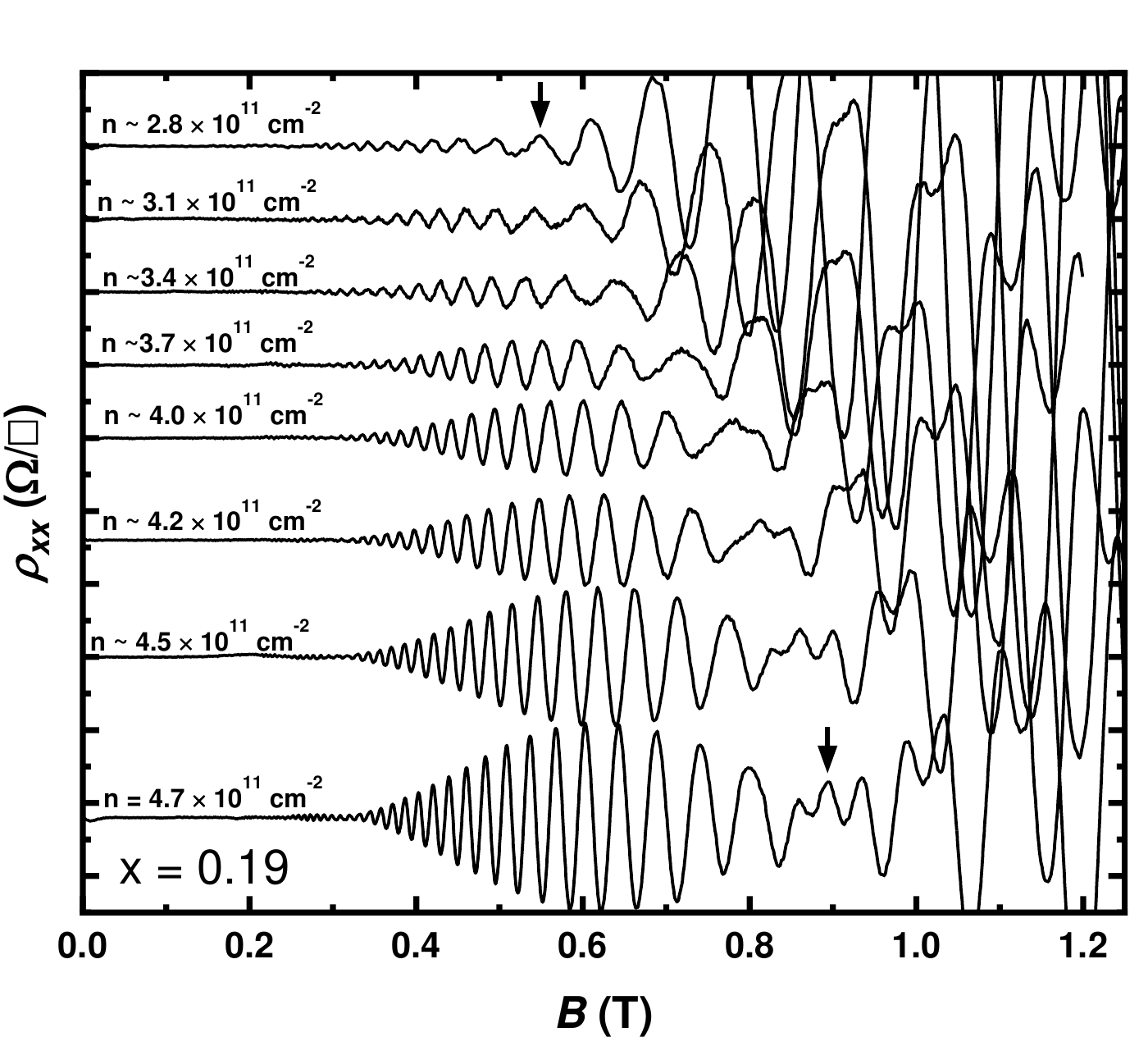}
    \caption{ SdH oscillations as a function of density for $x$ = 0.19. A shift of the 2nd node to higher magnetic field is observed as the density increases.}
    \label{Fig:fig5}
\end{figure}

The beating observed in low-field transport in our samples is attributed to zero field spin-splitting due to strong Rashba spin-orbit coupling. Zero-field spin-splitting ($\Delta_{so}$) is given by $2\alpha k_F$, where $k_F$ is the Fermi wave vector defined as $k_F=\sqrt{2 \pi n}$ and $\alpha$ is the Rashba coupling parameter. The presence of SOC results in beating in the SdH oscillations as seen in Fig. \ref{Fig:fig5}. Beating occurs because two occupied spin bands have two slightly different densities of electrons, giving rise to two closely spaced frequencies of SdH oscillations. This allows extraction of $\alpha$ from oscillatory low-field transport \cite{Das1989,Engels1997,Nitta1997,Heida1998,Schapers1998,Hatke2017}. Fourier analysis of longitudinal magnetoresistance facilitates extraction of the Rashba parameter $\alpha$ as a function of 2DEG density and arsenic mole fraction in our devices.   

We assume the Rashba coupling is the major contribution to SOC in our devices. Due to the lack of microscopic inversion symmetry in zinc-blende crystals, the $k^3$ Dresselhaus coupling also contributes to SOC in zinc-blende InSbAs quantum wells \cite{Nishizako2010,Kallaher2010}. However, Rashba coupling is known to dominate in highly asymmetric quantum wells such as those studied here \cite{Khodaparast2004,Gilbertson}. The increased electrostatic asymmetry with increased gate voltage is confirmed by self-consistent calculation of the band diagram as shown in Fig. \ref{Fig:fig3}b.

The Rashba parameter depends on the electric field in the quantum well and in the top and bottom barriers when the wavefunction has finite value in the barriers. The conduction band discontinuities at the top and bottom interfaces also contribute when the wavefunction has differing amplitude at the top and bottom interfaces \cite{Engels1997,Schapers1998}.

\begin{equation}\label{Eq1}
 \alpha \propto a_0 <E_{QW}> + b'(<E_u>+<E_l>)-b(|\Psi_u|^2-|\Psi_l|^2)\\
\end{equation}

The parametric dependence of $\alpha$ is shown in Eq. \ref{Eq1}, where the first term is the expectation value of the electric field in the quantum well weighted by $a_0$, which is a material specific parameter determined by interband matrix elements and energy differences \cite{winkler2003spin}. $a_0$ is expected to increase with increasing arsenic mole fraction in the InSbAs ternaries we study. The second term contains the expectation value of the electric field in the upper and lower barriers weighted by $b'$ which characterizes Rashba coupling in the barriers. The third term accounts for the contributions from the wavefunction at the upper ($|\Psi_u|^2$) and lower interface ($|\Psi_l|^2$), where $b$ parameterizes band discontinuity at the interfaces.  

The Rashba parameter is determined once $\Delta n$, the density difference between the two electron spin bands, is known \cite{Engels1997}. This is accomplished by taking a fast Fourier transform of the low-field longitudinal resistance, which yields the frequencies of the two oscillations, $f_1$ and $f_2$. The Rashba parameter is given by Eq. \ref{Eq2} \cite{Engels1997}.
\begin{equation}\label{Eq2}
\alpha = \frac{\Delta n \hbar^2}{m^*} \sqrt{\frac{\pi}{2(n_{tot}-\Delta n)}}\\
\end{equation}
In Eq. \ref{Eq2}, $n_{tot}=n_1+n_2$, $\Delta n = n_1-n_2$ and $m^*$ is the effective mass. We use the effective mass values measured for our samples as shown in Table \ref{table:info}.

\begin{figure}[ht]
    \includegraphics[width=\columnwidth]{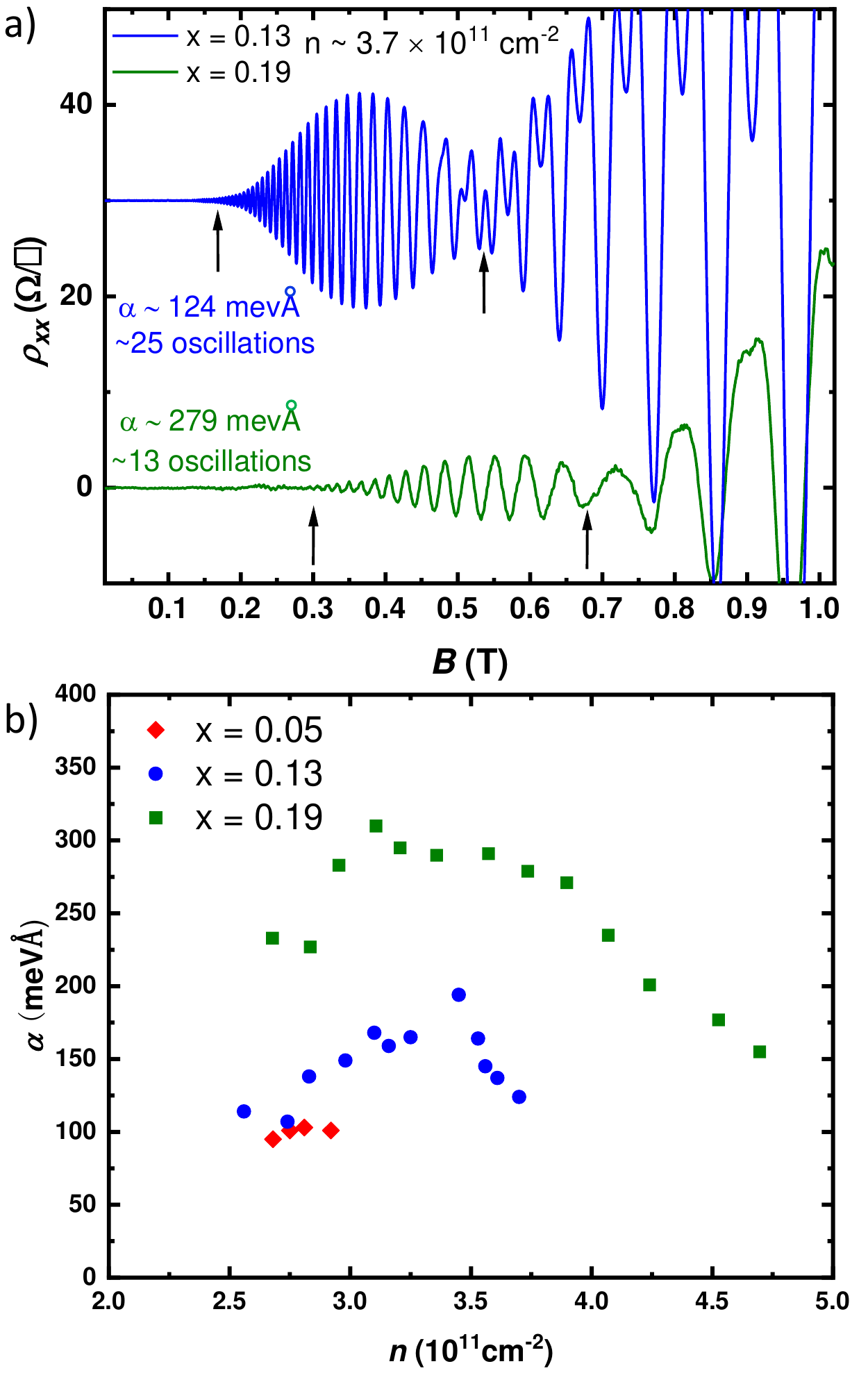}
    \caption{ a) Direct comparison of beating in SdH oscillations with $x$ = 0.13 and $x$ = 0.19 at $n = 3.7$ $ \times $ $10^{11}$ cm${^{-2}}$. Comparison at the same density allows for comparison at similar electric field. The number of oscillations between two nodes is inversely proportional to the value of $\Delta_{so}$. For $x$ = 0.13 we observe a larger number of oscillations compared to $x$ = 0.19, confirming a lower value of zero field spin splitting energy. b) Extracted Rashba parameter as a function of 2DEG density for all samples.}
    \label{Fig:fig6}
\end{figure}

The increase of the Rashba coupling with larger arsenic mole fraction is evident in the raw magneto-resistance data plotted in Fig. \ref{Fig:fig6}a. The number of oscillations between two nodes changes from approximately 25 to 13  from $x$ = 0.13 to $x$ = 0.19. Since the number of oscillations is inversely proportional to $\Delta_{so}$, this indicates a higher $\alpha$ for the higher arsenic mole fraction sample \cite{Nitta1997,Luo1988}. The evolution of the Rashba parameter with density for all arsenic mole fractions is depicted in Fig. \ref{Fig:fig6}b. A noticeable increase of the Rashba coupling with increased arsenic mole fraction is evident. If we compare the Rashba coupling extracted at the same density, $n \approx$ 2.9 $\times$ 10$^{11}$ cm${^{-2}}$, the values for Sample A to C are respectively $\alpha \approx$ 100, 150, 280 meV \text{\AA}. Because the samples have the same top and bottom barriers, nearly equal electric field in the quantum well and matching electron distributions (as inferred from simulations), the increase in SOC may be attributed to an increase in the material specific Rashba coupling parameter $a_0$ with higher arsenic concentration. This is consistent with predictions \cite{Berolo,Bouarissa,Winkler2016, mayer2020superconducting}. 

For Sample B and Sample C, the Rashba coupling increases from $n \approx$ 2.5 and 3.5 $\times$ $  10^{11}$ cm$^{-2}$. This enhancement is consistent with simulations, which indicate an increase in electric field and wavefunction asymmetry towards the top barrier, as also seen in Fig. \ref{Fig:fig2}b. However, for both Samples B and C, the Rashba parameter decreases after $n$ $\approx$ 3.5 $\times$ $ 10^{11}$ cm$^{-2}$. It is worth noting that the decrease in $\alpha$ occurs at approximately the same density at which the second electric subband begins to be occupied in the numerical simulations, although contribution from a second subband to the magnetotransport is not apparent in the data. From the simulations, the second subband wavefunction penetrates significantly more into the top barrier than the first subband. A possible explanation for the decrease in $\alpha$ with higher density is the increased propensity of electrons to reside in the top barrier, which has lower intrinsic SOC than the quantum well. However, more investigation is needed to quantitatively understand this trend.
 
\section{\label{conc}Conclusions}
In this work we have detailed the low temperature transport properties of a series of 2DEGs confined in InSbAs quantum wells with different arsenic mole fractions. We extracted an alloy scattering rate of 1/$\tau$ = 45 ns$^{-1}$ per $\%$ As, determined the dependence of mobility on 2DEG density and arsenic mole fraction, and detailed the evolution of dominant scattering mechanisms. Analysis of SdH oscillations enabled extraction of the Rashba coupling strength over a wide range of 2DEG density for all mole fractions. At comparable density and nearly equal electric field, we observed an increase of $\alpha$ with arsenic mole fraction. For $x$ = 0.19 $\alpha$ is $\approx$ 300 meV$ \text{\AA}$, one of the highest values reported for III-V heterostructures. These results agree with theoretical predictions of enhancement of SOC with arsenic mole fraction \cite{Winkler2016,Berolo,Bouarissa} and suggest that InSb$_{1-x}$As$_{x}$ 2DEGs are a promising platform for exploration of topological superconductivity by incorporating these ternary quantum wells in a shallow structure coupled to an s-wave superconductor. 

\nocite{*}

\bigskip
{$\dagger$} Author to whom correspondence should be addressed: mmanfra@purdue.edu

\bigskip
We thank Dr. James Nakamura for suggestions for manuscript improvement and Georg Winkler, Sam Teicher and Fabiano Corsetti for fruitful discussion. This work was supported by Microsoft Quantum. 

\bigskip
The data that support the findings of this study are available from the corresponding author upon reasonable request.

\bigskip
The authors have no conflicts to disclose.

\bibliography{apssamp}% Produces the bibliography via BibTeX.

\end{document}